\documentclass[aps,superscriptaddress,twocolumn]{revtex4}

\usepackage{amsmath, amsthm, amsfonts, amssymb}
\usepackage{arydshln}
\usepackage{array}
\usepackage{graphicx}
\usepackage{rotating}
\usepackage{tikz}
\usetikzlibrary{shapes.geometric, patterns}
\usetikzlibrary{decorations.pathreplacing, angles, quotes}
\usetikzlibrary{arrows, decorations.markings}
\usetikzlibrary{plotmarks}
\tikzset{
    buffer/.style={
        draw,
        regular polygon
    }
}

\theoremstyle{plain}

\theoremstyle{definition}

\def\F{F_\nu(1,2)}

\begin{document}

\title{Spanning Trees of Recursive Scale-Free Graphs}
\author{C. Tyler Diggans}

\affiliation{Clarkson Center for Complex Systems Science, Clarkson University, Potsdam, NY 13699} 
\affiliation{Department of Physics, Clarkson University, Potsdam, NY 13699} 
\affiliation{Air Force Research Laboratory: Information Directorate, Rome, NY 13441}
\author{Erik M. Bollt}
\affiliation{Clarkson Center for Complex Systems Science, Clarkson University, Potsdam, NY 13699} 
\affiliation{Department of Electrical and Computer Engineering, Clarkson University, Potsdam, NY 13699} 
\author{Daniel ben-Avraham} 
\affiliation{Clarkson Center for Complex Systems Science, Clarkson University, Potsdam, NY 13699} 
\affiliation{Department of Physics, Clarkson University, Potsdam, NY 13699} 

\begin{abstract}
We present a link-by-link rule-based method for constructing all members of the ensemble of spanning trees for any recursively generated, finitely articulated graph, such as the DGM net.  The recursions allow for many large-scale properties of the ensemble of spanning trees to be analytically solved exactly.  We show how a judicious application of the prescribed growth rules selects for certain subsets of the spanning trees with particular desired properties (small-world, extended diameter, degree distribution, etc.), and thus approximates and/or provides solutions to several optimization problems on undirected and unweighted networks.  The analysis of spanning trees enhances the usefulness of recursive graphs as sophisticated models for everyday life complex networks.
\end{abstract}
\maketitle

\section{Introduction}
Recursive, finitely articulated models of complex networks mimic many key properties of real-world networks.  At the same time, these models are amenable to {\em exact} analysis, providing important insights into the nature of everyday life complex networks.  In this article, we revisit the subject of spanning trees for recursive scale-free nets, and provide a method for generating all spanning trees, enabling a more complete analysis of this important set of subgraphs.  

The link-based method presented here is applicable to any recursive generative network model that produces finitely articulated graphs, but we focus on the well-known DGM net~\cite{DGM} as a representative example of this class.  This particular family of networks was introduced in~\cite{DGM} as a \textit{Pseudo-fractal Scale-free Web} (and was referred to as the PSW in subsequent literature), but is equivalent to the $(u,v)$-flower graph~\cite{Rozenfeld06} when $u=1$ and $v=2$.  We use our notation, $\F$, to denote the $\nu$-th generation of this network~\cite{Rozenfeld06}, for which there are two equivalent ways to construct it recursively: the link-by-link approach and the hub-pasting approach.  Both recursive growth processes create the same set of small-world complex networks with hierarchical structure, a scale-free degree distribution with power law exponent $\gamma\approx2.57$, and a high clustering coefficient of $C=4/5$, matching several characteristic benchmarks of real-life complex nets.  The recursive nature and the finite articulation have enabled exact analytical studies of the statistics of cycles~\cite{Rozenfeld04}, diffusion~\cite{Bollt05}, percolation~\cite{Rozenfeld07}, spectral properties~\cite{Xie16}, and minimum dominating sets~\cite{Shan17} for this network, shedding light on the analogous properties in real-life complex nets.  Previous work has exploited the hub-pasting approach of constructing this and similar models to enumerate the spanning trees (and calculate the subsequent \textit{tree entropy}~\cite{Lyons05} of the ensembles)~\cite{Zhang10_PSW, Zhang11_trees, Zhang14_trees, Sun16_trees}, but additional results related to spanning trees of these networks are now obtained by considering a link-by-link approach.  

After providing a brief description of both methods of constructing the original $\F$ in section~\ref{Fuv}, a few new results on the spanning trees of these graphs will be provided from the hub-pasting viewpoint in section~\ref{hubpasting}, before turning to the link-based approach.  In section~\ref{linkbylink}, we provide an explicit process for efficiently generating the whole ensemble of spanning trees. We use the same recursive link-by-link process that creates the original network, except that the links are assigned an $R$ or $G$, which stand for ``real" or ``ghost" respectively.  The term ghost indicates that a link is part of the original graph, but not part of the spanning tree, and we represent these links by dashed lines in all the figures. 

Using the link-based approach, we first verify previous results for the enumeration of the labeled spanning trees of $\F$ in a new simplified way and calculate the tree entropy of the general $(u,v)$-flower for comparison with previous results.  We then show that this approach finds added utility in solving optimization problems on undirected and unweighted networks of this type.  While the well known Minimum Spanning Tree (MST) problem on weighted networks has garnered a lot of attention~\cite{Graham85}, because of its relevance to the famed {\em traveling salesman} problem~\cite{Held70,Magnanti84} and its useful applications in data science~\cite{Shapiro95,Xu02,Varma04,He08,Galluccio12,Maeng12,Rainbolt17}; the MST is not clearly defined for unweighted networks.  Common generalizations do exist, however, for the undirected and unweighted case including the dense, sparse, and minimum routing cost spanning tree problems, as well as many other lesser studied variations~\cite{Wu04,Dobrynin01,Kim04,Salamon08,Cerulli09,Silva14, Szekely16, Li17, Silvestri17, Merabet18, Ozen20,  Bozovic21}.  The solutions to these problems are relevant to an ever widening diversity of applications from optical network design~\cite{Gargano02,Marin15} to networked oscillator synchronization~\cite{Diggans21}.  We show that while these combinatorial problems are often NP-hard in general~\cite{Gargano02,Salamon08,Silvestri17}, for recursive, finitely articulated networks, such optimal spanning trees (and indeed the whole ensemble of spanning trees) can now be created alongside the original network with minimal overhead.  We conclude this article by describing a process for selecting such optimal spanning trees through imposing restrictions on the application of the link-based rules, and provide an interesting open problem that we hope this work sheds light on.

\section{Constructing $F_\nu(u,v)$}
\label{Fuv}
\begin{figure}[ht!]
\begin{tabular}{cc}
\includegraphics[width=3.25cm]{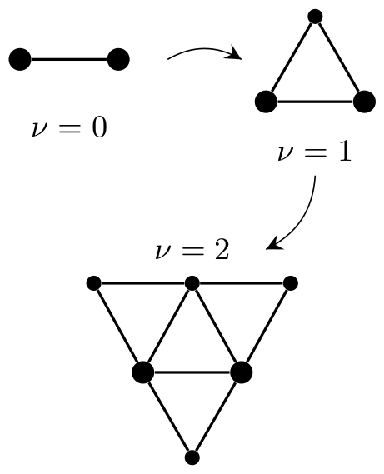}&
\includegraphics[width=4.25cm]{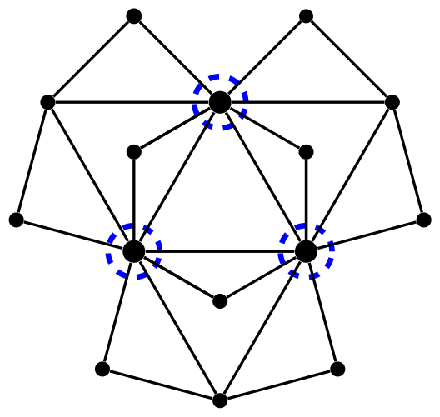} \\
(a) & (b)\\
\end{tabular}
\caption{Two equivalent methods of constructing $\F$ from $F_{\nu-1}(1,2)$: (a) the first three generations of $\F$ are obtained by adding a path of length $2$ in parallel to each link from the previous generation and (b) the construction of $F_3(1,2)$ is obtained by identifying the hub vertices of $u+v=3$ copies of $F_2(1,2)$; the three nodes resulting from the hub identifications (the new hubs) are indicated by dashed circles for clarity.\label{UV}}
\end{figure}

Starting from an initial seed graph of the complete graph on two vertices, i.e. $K_2$, in each generation of the growth process, we replace each link from the previous generation with a pair of paths of length $u$ and $v$, in parallel, between the nodes of the original link. For the case of $u=1$ and $v=2$, this is equivalent to just adding a new path of length two in parallel to the existing link as shown in Fig.~\ref{UV} (a), which illustrates the first three generations of $\F$.  

For this case, it follows that for each link from the previous generation, one new node is added to the network, while $w=u+v=3$ links replace each old link in the new generation. Thus, the size (total number of links), $M_\nu$, and order (total number of nodes), $N_\nu$, of the $\nu$-th generation are given by $M_\nu = w^\nu = 3^\nu$ and
\begin{equation}
N_\nu=\frac{w-2}{w-1} w^\nu + \frac{w}{w-1}=\frac{1}{2} 3^\nu + \frac{3}{2},
\label{Nnu}
\end{equation}
where detailed derivations through recursion relations are provided elsewhere~\cite{DGM,Rozenfeld06,Diggans20}.  

Alternately, one can construct $\F$ using a hub-pasting approach by taking $w=u+v=3$ copies of $F_{\nu-1}(1,2)$ and identifying one hub vertex (those of highest degree) on each copy with another hub vertex on each other copy resulting in $\F$. This process is shown in Fig.~\ref{UV} (b) for the creation of $F_3(1,2)$ by hub-pasting three copies of $F_2(1,2)$, where the identified hub nodes are indicated by dashed circles.  

Prior to exploring the main contribution of this work within the context of the link-by-link approach, we first mention two points that have not previously been discussed, but are best described within the hub-pasting paradigm.

\section{Hub-pasting Construction and the Typical Diameter}
\label{hubpasting}
For generating spanning trees of $\F$, one can begin by selecting (with replacement) three members from the ensemble of trees from generation $\nu-1$ and identifying them at the vertices that would be considered hubs of $F_{\nu-1}(1,2)$ as shown in Fig~\ref{Constructions} (a).  We will use the italicized version of the word \textit{hub} to indicate these hub nodes of the original network when in reference to a spanning tree, whether those nodes in fact have the largest degree in the tree or not.  This pasting process is done explicitly by temporarily adding a link between the \textit{hub} vertices from adjacent spanning trees (chosen from the previous generation), and then contracting the graph along these temporary links to identify the three pairs of \textit{hub} nodes into the \textit{hubs} for the new generation. Due to the connectedness of the $3$ individual spanning trees, the identification process will form a single cycle through the new \textit{hub} nodes (shown in bold). To obtain a spanning tree of the new generation, remove one of the links from this cycle.  The ambiguity of this link removal step complicates the enumeration of spanning trees, which led to a more arduous accounting process used in previous work~\cite{Zhang10_PSW}.  However, this simplified description provides insight into the typical diameter of the spanning trees of $\F$.

In~\cite{Rozenfeld04}, the typical cycle length of $\F$ was shown to scale as $2^\nu$ in the thermodynamic limit of large graphs. This implies that the typical spanning tree of $\F$ contains a path having length on the order of $2^\nu$. Given that the maximum diameter of any tree is also of order $2^\nu$ (see section~\ref{Special}), then the typical spanning tree must have a diameter of order $2^\nu$.  It is interesting that while the $(1,2)$-flower is a small-world network, the diameter of the typical spanning tree of this network scales like $\sim 2^\nu\sim N^{\log_3 2}$ (where the order of $\F$ from Eq.~(\ref{Nnu}) is $N=\frac{1}{2}3^\nu+\frac{3}{2}\sim 3^\nu$) and is therefore \textit{not} small-world.  However, section~\ref{Special} shows how to select a small-world spanning tree from the (typically non small-world) ensemble.

Additionally, we note that the ``copy machine" algorithm outlined in Appendix~C of~\cite{Bollt05}, which is based on the hub-pasting approach, can be adapted to efficiently generate adjacency matrices of spanning trees of $\F$ in a manner similar to what was outlined for ensembles in~\cite{Diggans20}.

\section{Link-by-link Construction and Tree Entropy}
\label{linkbylink}
The \textit{link-by-link} approach to spanning tree construction will mirror the general link-by-link construction process of the recursive net, while keeping track of two different kinds of links: ``real" links ($R$) that belong to the spanning tree, and ``ghost" links ($G$) that belong to the original network, but not the spanning tree.

\begin{figure}[ht!]
\centering
\begin{tabular}{c}
\includegraphics[width=7cm]{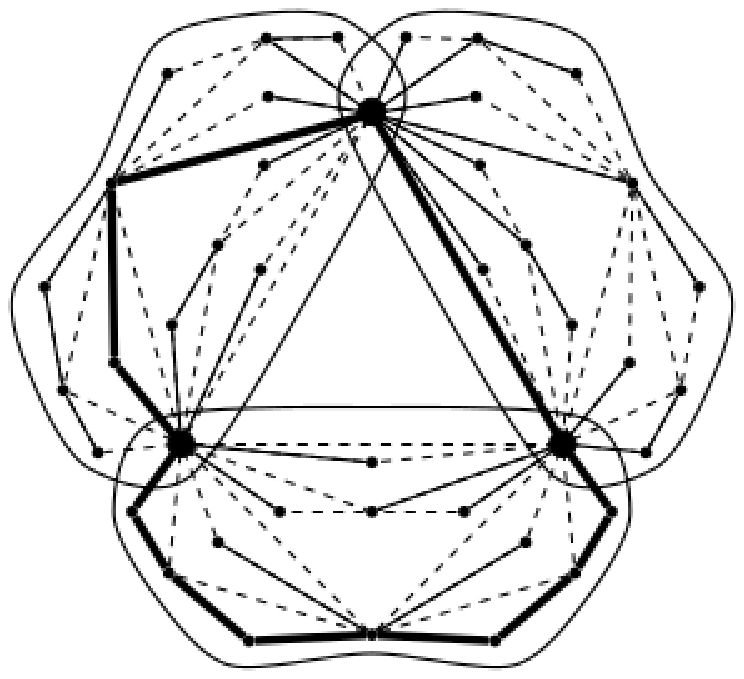}\\
(a)\\
\includegraphics[width=7.5cm]{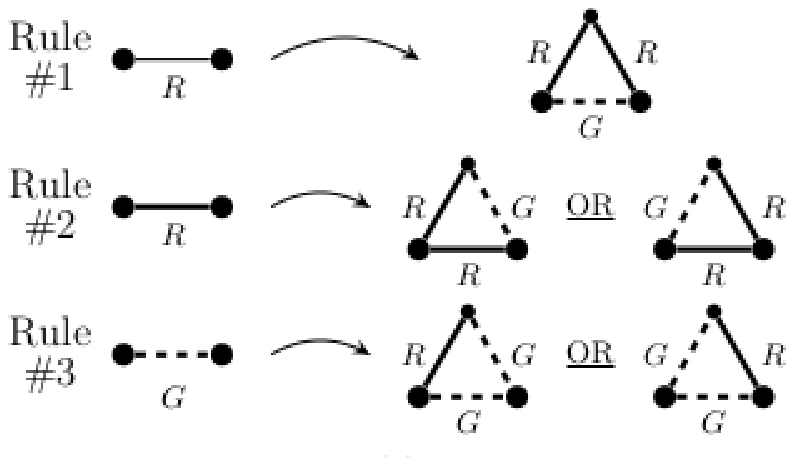}\\
(b)\\
\end{tabular}
\caption{As with the regular flower graph, there are two methods of constructing spanning trees of $(1,2)$-flower graphs: (a) the hub-pasting approach (here requiring an additional step) entails choosing $1+2=3$ elements from the previous generation of spanning trees and identifying pairs of \textit{hub} nodes as shown; this, however, results in a single cycle (bolded), which must be broken by removing one of its links. (b) The \textit{link-by-link} approach involves applying one of three rules to each link in each generation recursively. Each sequence of valid rule operations results in a unique spanning tree, allowing for the enumeration \textit{and generation} of all spanning trees in the ensemble.\label{Constructions}}
\end{figure}

For $\F$, in each generation, an $R$-link is either replaced by a $G$-link and two $R$-links (rule \# 1 in Fig.~\ref{Constructions} (b)), or  the $R$-link is replaced by an $R$-link and a path of one $R$-link and one $G$-link in either order (rule \# 2). Additionally, in the process going forward, a $G$-link is always replaced by a $G$-link and a path of one $R$-link and one $G$-link (rule \# 3).  Other recursive scale-free network models such as the generic $(u,v)$-flower graph~\cite{Rozenfeld06} or a similar construction described in~\cite{Zhang07} may require a larger set of rules with more cases to consider, making the subsequent analysis more tedious, but the process is, in essence, the same.

Denoting by $R_\nu$ ($G_\nu$) the number of real (ghost) links in generation $\nu$, the construction rules in Fig.~\ref{Constructions} imply
\begin{equation}
\label{RG}
\begin{aligned}
R_{\nu+1} &= 2 R_\nu + G_\nu, \qquad R_0=1\\
G_{\nu+1} &= R_\nu + 2 G_\nu, \qquad G_0=0,\\
\end{aligned}
\end{equation}
leading to $R_\nu=\frac{1}{2}\left(3^n+1\right)$ and $G_\nu=\frac{1}{2}\left(3^n-1\right)$. This is consistent with the fact that $R_\nu=N_\nu-1$ and $G_\nu+R_\nu=M_\nu$, where $M_\nu$ and $N_\nu$ are as defined in Eq.~(\ref{Nnu}).

Each sequence of link updatings results in a single unique spanning tree; moreover, each updating rule is uniquely reversible.  Finally, the reverse rules applied (in an appropriate manner) to an arbitrary spanning tree of $\F$ generate a spanning tree of $F_{\nu-1}(1,2)$ (i.e. no loops are created, and neither is the tree disconnected by the reverse rule application). It follows that there is a one-to-one correspondence between spanning trees and valid generating sequences of rules, and therefore, these rules generate the complete set of labeled spanning trees.

This approach enables an independent verification of the number of labeled spanning trees of generation $\nu$ for the DGM net~\cite{Zhang10_PSW}. Denote the number of spanning trees by $T_\nu$, and consider the recursion
\begin{equation}
\label{Tn}
T_{\nu+1} = \left(3^{R_\nu} \cdot 2^{G_\nu}\right) T_\nu,
\end{equation}
where the term in parenthesis accounts for the 3 (2) ways of updating $R$ ($G$) links. Together with the expressions for $R_\nu$ and $G_\nu$ and the initial condition of $T_0=1$, we get
\begin{equation*}
T_{\nu} = \left(\frac{3}{2}\right)^{\frac{\nu}{2}} 6^{\left(3^{\nu}-1\right)/4} .
\end{equation*}
In terms of the order of the graph, $N$,
\begin{equation*}
T_{\nu} = \left(\sqrt{\frac{3}{2}}\right)^{\log_3{(2N-3)}} 6^{\left(\frac{N}{2}-1\right)}\sim \frac{2^C}{6}~N^C~\sqrt{6}^N, 
\end{equation*}
where $C=(1-\log_3(2))/2$ is a constant. Although a different numbering convention for the generation was used here, this confirms the result found in~\cite{Zhang10_PSW}. 

For comparison, the total number of all labeled trees of order $N$ is $N^{N-2}$~\cite{Cayley1889}. Amusingly, this functional dependence of $T_\nu$ with $N$ is closer to that of \textit{unlabeled} trees of order $N$~\cite{Otter48}.

For further comparison, this approach can be adapted to the general $(u,v)$-flower by letting $w=u+v$. The recursion for the enumeration of all labelled spanning trees [analogous to Eq.~(\ref{Tn})] is then: $T_{\nu+1} = w^{R_\nu} \cdot (uv)^{G_\nu} T_{\nu}$ with $T_0=1$, since there are $w$ ways that any $R$-link can be propagated (the spanning trees of a $w$ cycle) and $uv$ ways that any $G$-link can be propagated (one link in each path having lengths $u$ and $v$ must be a ghost link).  A recursion similar to Eq.~(\ref{RG}) then gives $R_\nu=\frac{w-2}{w-1}w^\nu+\frac{1}{w-1}$ and $G_\nu=\frac{1}{w-1}w^\nu-\frac{1}{w-1}$, which leads to the closed form solution
\begin{equation*}
T_{\nu} = w^{\frac{w-2}{w-1}\left(\frac{w^\nu-1}{w-1}+\frac{\nu}{w-2}\right)}\cdot (uv)^{\frac{1}{w-1}\left(\frac{w^\nu-1}{w-1}-\nu\right)}.
\end{equation*}
This gives the tree entropy~\cite{Lyons05} of the $(u,v)$-flower as
\begin{equation*}
\begin{aligned}
\lim_{N\rightarrow \infty}\frac{\ln{T_\nu}}{N}&=\frac{ln{(w)}}{w-1}+\frac{\ln{(uv)}}{(w-1)(w-2)}, 
\end{aligned}
\end{equation*}
So, while the $(1,2)$-flower indeed has a small tree entropy when compared with other networks having an average degree of $\langle k\rangle=4$ as described in~\cite{Zhang10_PSW}, it is also true that larger values of $w$ will result in smaller tree entropies tending toward zero in the limit of large $w$; although the average degrees of flower graphs will tend toward $\langle k\rangle=2$ as $w\rightarrow \infty$.  Finally, we note that the tree entropy for the $w$-cycle is simply $\frac{\ln{(w)}}{w}$, meaning the tree entropy of the flower graphs are increased by an additional term that is quadratic in $\frac{1}{w}$ due to the overlapping $w$-cycles that define them.

\section{Spanning Tree Subsets having optimal characteristics}
\label{Special}
In many applications of spanning trees, one is not concerned with simply the number of trees, but one seeks an optimal spanning tree for some purpose, e.g., minimal branching vertex spanning tree (MBVST), dense spanning tree (DST), etc.  Within the process outlined, due to the finite articulation of the models being addressed, large-scale properties can be defined in terms of recursion relations allowing such optimal trees to be selected from various subsets of the ensemble of spanning trees that share certain properties. This selection is done by restricting how the rules are applied, which alters the recursions that define these properties. 

To illustrate, consider the maximum diameter of any spanning tree from the ensemble in generation $\nu$, which we denote by $s_\nu^{\textrm{max}}$. If rule \# 1 is applied to all $R$ links, then for each $R$ link in the path that defined the diameter of the previous generation, we will have two $R$ links in the path that contribute to the diameter of the new generation.  At least for generation $\nu>2$, there will also be two additional $G$ links that will enable the diameter to be extended further, meaning we have the recursion $s_\nu^{\textrm{max}} = 2 s_{\nu-1}^{\textrm{max}} + 2$, with initial condition $s_2^{\textrm{max}}=5$, leading to the formula $s_\nu^{\textrm{max}}=7\cdot2^{\nu-2}-2$.  So, applying rule \# 1 to all $R$-links (excluding rule \# 2 entirely) leads to the diameter scaling $\sim 2^\nu$, and in fact, the application of any finite proportion of rule \# 1 to $R$-links will result in a \textit{non} small-world spanning tree of these small-world networks.  

\begin{figure}
\centering
\begin{tabular}{c}
\includegraphics[width=6.5cm]{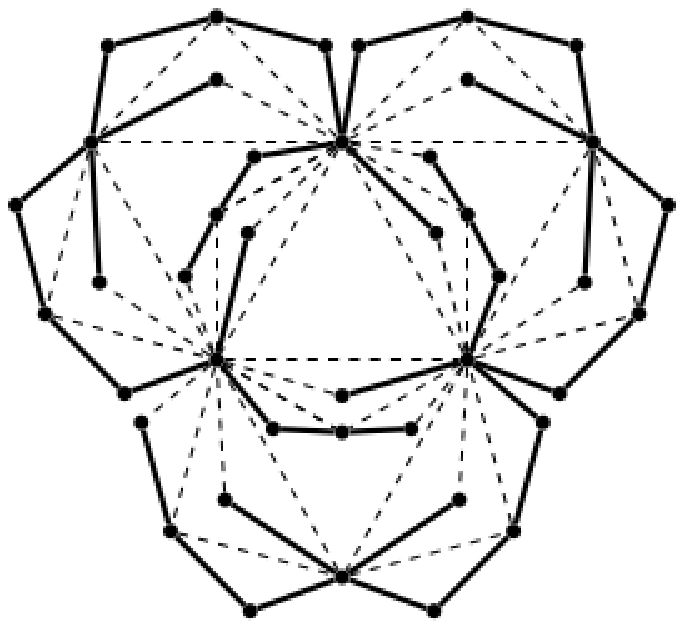}\\
(a)\\
\includegraphics[width=6.5cm]{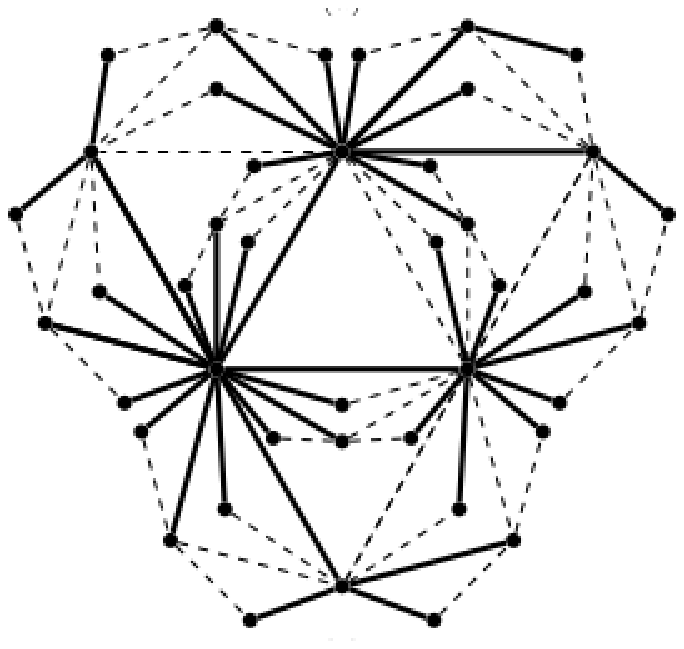}\\
(b)\\
\includegraphics[width=6.5cm]{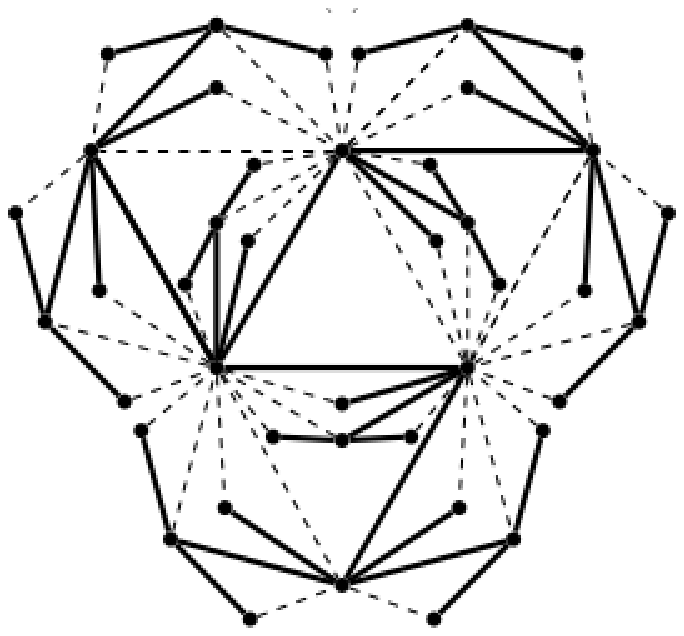}\\
(c)\\
\end{tabular}
\caption{Representative examples of subsets of spanning trees that can be selected from within the set of spanning trees of $\F$: (a) using rule \#'s 1 and 3 only, where the ambiguity of rule \# 3 is reduced by maximizing the diameter when possible (breaking ties randomly) resulting in a solution to the MBVST problem; (b) using rules \#'s 2 and 3 only with $p=1$ resulting in a solution to the DST problem; and (c) using rule \#'s 2 and 3 only with $p=0$ leads to a balanced small world tree with a more homogeneous degree distribution that is a candidate for a solution to the $k$-MBVST problem.\label{Examples}}
\end{figure}

These maximal diameter spanning trees are solutions to both the minimal branching vertex and the sparse spanning tree problems (MBVST and SST respectively)~\cite{Gargano02,Cerulli09,Ozen20, Bozovic21}.  So by applying rule \# 1 to all $R$-links and reducing the ambiguity of rule \#3 by choosing the $R$ link to extend the diameter whenever possible (e.g. choosing the $R$ link to be adjacent to nodes with smaller degree), a member of the subset of spanning trees with maximal diameter is obtained, see Fig.~\ref{Examples} (a) for one of the subset of eight such spanning trees of $F_4(1,2)$ having $s_4^{\textrm{max}}=26$ that are in fact solutions to the MVBST and  SST problems.  

Contrast this spanning tree with a spanning tree that is created by never applying rule \# 1, which is a small-world spanning tree since the diameter then scales as the logarithm of the order.  In the class of trees where only rule \#'s 2 and 3 are applied, if one chooses $R$-links to be adjacent to the largest degree node (breaking ties randomly), then one obtains a member of the subset of spanning trees with the smallest diameter, i.e. a solution to the dense spanning tree problem, which has minimal Wiener index~\cite{Wiener47,Szekely16, Ozen20} (see Fig.~\ref{Examples} (b)). In particular, we obtain a spanning tree having the minimum diameter, defined by the recursion $s_{\nu+1}^{\textrm{min}} = s_{\nu}^{\textrm{min}} + 1$, since the ambiguity of rule \# 2 prevents any increase in the span, but there is one ghost link in each generation that must add to the span.  The initial condition of $s_{1}^{\textrm{min}}=2$ leads to $s_{\nu}^{\textrm{min}} = \nu + 1$, meaning in terms of the order of the graph, we have $s_{\nu}^{\textrm{min}} \sim log_3(N)$ in the limit of large graphs.

Letting a parameter $p$ control the proportion of instances (when applying rule \#'s 2 and 3) where the $R$-link is attached to the largest degree node, the tree in Fig.~\ref{Examples} (b) is a representative of the set created using $p=1$; and the tree in Fig.~\ref{Examples} (c) is representative of using $p=0$.  The parameter $p$, then alters the degree distribution while retaining the small-world property, and values of $p\in(0,1)$ then define probability distributions over this subset of trees having small Wiener indices, providing a useful tool for approaching the Minimal $k$-Branching Vertex problem (k-MBVST)~\cite{Merabet18,Bozovic21}.

\section{Conclusion}
In summary, a general approach for obtaining spanning trees of recursive, finitely articulated graphs has been presented through the example of the DGM net, and basic properties of the ensemble of all spanning trees, such as the typical/maximum diameter and the number of labeled trees are obtained exactly.  Additionally, one can now select a spanning tree from subsets of the ensemble of spanning trees that share desirable properties such as small-world features, certain degree distributions, and maximal/minimal diameter, effectively solving many optimal spanning tree problems and approximating solutions to others on such finitely articulated models. 

Perhaps the most interesting related open question is to find the number of \textit{unlabelled} spanning trees of recursive finitely articulated graphs. While this is, in general, an intractable problem, we hope that the recursive properties of these graphs might provide a convenient foothold to approach such a problem.
\section{Acknowledgments}
CTD has recieved funding from the Air Force Office of Scientific Research (20RICOR010)

EB has received funding from the Army Research Office (N68164-EG) and also DARPA

\bibliographystyle{unsrt}
\bibliography{STSFG_Re_arXiv.bib}

\end{document}